\documentclass[12pt,a4paper]{article}%
\usepackage{amsmath}
\usepackage{amsfonts}
\usepackage{amssymb}
\usepackage{graphicx}%
\providecommand{\U}[1]{\protect\rule{.1in}{.1in}}

\begin{document}
\begin{titlepage}
\title{Line-Integral Representations of the Diffraction of Scalar Fields}
\date{}
\author{Yi-Chuan Lu \and \textit{Department of Physics, National Taiwan University,}\\ \textit{Taipei 10617, Taiwan, R.O.C}}
\maketitle
\begin{abstract}
Traditionally, the diffraction of a scalar wave satisfying Helmholtz equation
through an aperture on an otherwise black screen can be solved approximately
by Kirchhoff's integral over the aperture. Rubinowicz, on the other hand, was
able to split the solution into two parts: one is the geometrical part
that appears only in the geometrical illuminated region, and the other
representing the reflected wave is a line-integral along the edge of the
aperture. However, this decomposition is not entirely satisfactory in
the sense that the two separated fields are discontinuous at the boundary of
the illuminated region. Also, the functional form of the line-integral is not
what one would expect an ordinary reflection wave should be due to some
confusing factors in the integrand. Finally, the boundary conditions on the
screen imposed by Kirchhoff's approximation are mathematically inconsistent,
and therefore, rigorously, this decomposition formulation must be
slightly modified by taking into account the correct B.C.s.\\
In this thesis, we use the consistent boundary conditions to derive a slightly
different decomposition formula which shows that the behavior of
the diffracted wave at the edge is exactly just like an ordinary
reflection---realizing the conjecture of Thomas Young in the 18$^{th}$ century. We
also derived another decomposition formula which avoids mathematical
discontinuity encountered by Rubinowicz. In the last section we demonstrate that our
solution is consistent with that obtained by Sommerfeld in the
rigorous 2-D plane-wave diffraction problem, so our formulation in this sense
may describe more accurately the behavior of diffracted wave near the edge of the
aperture than Kirchhoff's formula.
\end{abstract}
\end{titlepage}

\section{\bigskip Introduction}

\subsection{Diffraction Integral Formulae}

The diffraction of a scalar wave $\psi$ satisfying the Helmholtz equation%
\[
\left(  \nabla^{2}+k^{2}\right)  \psi=0
\]
has the solution (at the field point $\vec{r}_{f}$):%
\begin{equation}
\psi\left(  \vec{r}_{f}\right)  =-%
{\displaystyle\oint_{\mathcal{S}}}
\left(  \psi\frac{\partial G}{\partial n}-G\frac{\partial\psi}{\partial
n}\right)  da \label{psi in general}%
\end{equation}
where the integral is performed on a closed surface $\mathcal{S}$ which does
not enclose any source of $\psi,$ and $\hat{n}$ is the outward normal of
$\mathcal{S}.$ Here, $G$ is the Green's function satisfying%
\[
\left(  \nabla^{2}+k^{2}\right)  G=-\delta^{\left(  3\right)  }\left(  \vec
{r}-\vec{r}_{f}\right)  .
\]

For the problem of diffraction through an aperture on an infinite screen, one
usually defines $\mathcal{S}$ to be the union of the aperture, the screen, and
the infinity. If the aperture is finite and the screen is opaque, one expects
that $\psi$ decreases fast to zero at infinity, and thus one only has to
evaluate the integral on the aperture and the screen. Kirchhoff further
assumed the boundary conditions%
\begin{equation}
\left\{
\begin{array}
[c]{ll}%
\psi=\psi_{s}\text{ and }\dfrac{\partial\psi}{\partial n}=\dfrac{\partial
\psi_{s}}{\partial n} & \text{on the aperture,}\\
\psi=0\text{ and }\dfrac{\partial\psi}{\partial n}=0 & \text{on the screen,}%
\end{array}
\right.  \label{BC inconsistent}%
\end{equation}
where $\psi_{s}$ is the unperturbed source field. Also, he Kirchhoff the
Green's function%
\[
G\equiv G_{K}\equiv\frac{1}{4\pi}\frac{e^{ik\left\Vert \vec{r}-\vec{r}%
_{f}\right\Vert }}{\left\Vert \vec{r}-\vec{r}_{f}\right\Vert }.
\]
With these assumptions, Equation (\ref{psi in general}) can be reduced to%
\begin{equation}
\psi\left(  \vec{r}_{f}\right)  =-\int_{\text{aperture}}\left(  \psi_{s}%
\frac{\partial G_{K}}{\partial n}-G_{K}\frac{\partial\psi_{s}}{\partial
n}\right)  da. \label{psi Kirchhoff}%
\end{equation}

However, the boundary conditions imposed by Kirchhoff is mathematically
inconsistent, thought it gives good approximations near the axis at far field
zone. Sommerfeld, on the other hand, suggested another consistent boundary
conditions%
\begin{equation}
\left\{
\begin{array}
[c]{ll}%
\psi=\psi_{s}, & \text{on the aperture,}\\
\psi=0, & \text{on the screen,}%
\end{array}
\right.  \label{BC consistent}%
\end{equation}
and adopted Green's function of Dirichlet type which vanishes on the aperture
and the screen. For example, for a planar screen,
\[
G\equiv G_{D}\equiv G_{K}-G_{K}^{\ast}\equiv\frac{1}{4\pi}\frac
{e^{ik\left\Vert \vec{r}-\vec{r}_{f}\right\Vert }}{\left\Vert \vec{r}-\vec
{r}_{f}\right\Vert }-\frac{1}{4\pi}\frac{e^{ik\left\Vert \vec{r}-\vec{r}%
_{f}^{\ast}\right\Vert }}{\left\Vert \vec{r}-\vec{r}_{f}^{\ast}\right\Vert },
\]
where $\vec{r}_{f}^{\ast}$ is the mirror image of $\vec{r}_{f}$ with respect
to the screen. With these modifications, Equation (\ref{psi in general}) is
reduced to%
\begin{equation}
\psi\left(  \vec{r}_{f}\right)  =-\int_{\text{aperture}}\psi_{s}\frac{\partial
G_{D}}{\partial n}da. \label{psi Sommerfeld}%
\end{equation}

\subsection{Maggi-Rubinowicz' Decomposition}

With the Kirchhoff integral formula (Equation (\ref{psi Kirchhoff})),
Rubinowicz was able to decompose the field $\psi\left(  \vec{r}_{f}\right)  $
into two parts \cite{Rubinowicz Original}: one that appears only in the
ordinary geometrical illuminated region is the source field evaluated at the
field point $\vec{r}_{f}$; the other one is a line integral along the edge of
the aperture:%
\begin{equation}
\psi\left(  \vec{r}_{f}\right)  =\left\{
\begin{array}
[c]{cl}%
\psi_{s}\left(  \vec{r}_{f}\right)  & ,\vec{r}_{f}\in\text{illuminated
region}\\
0 & ,\text{otherwise}%
\end{array}
\right.  -\frac{1}{4\pi}%
{\displaystyle\oint\nolimits_{\text{edge}}}
\psi_{s}\left(  \vec{r}\right)  \frac{e^{ik\rho_{f}}}{\rho_{f}}\left(
\frac{\hat{\rho}_{s}\times\hat{\rho}_{f}}{1+\hat{\rho}_{s}\cdot\hat{\rho}_{f}%
}\right)  \cdot d\vec{l}. \label{Psi Rubinowicz}%
\end{equation}

This formula applies to two special cases. The first one is the
plane-wave-incidence case in which the source field is a plane wave: $\psi
_{s}\left(  \vec{r}\right)  =e^{ik\rho}$, where $\rho$ is the distance
measured from $\vec{r}$ to a constant phase plane of the incident field. If
$\vec{r}$ happens to lie on the edge of the aperture, then we denote $\rho$ by
$\rho_{s}.$ Also, we denote $\hat{\rho}_{s}$ to be the unit vector in the
direction of propagation of the incident field, as shown in Figure (1). The
illuminated region is an oblique cylinder, as predicted by geometrical optics.
Inside the line integral, $\rho_{f}\equiv\left\Vert \vec{r}-\vec{r}%
_{f}\right\Vert $ is the distance from $\vec{r}_{f}$ to the edge $\vec{r}$,
and $\hat{\rho}_{f}$ is the unit vector of $\vec{r}-\vec{r}_{f}$.

The second one is called the point-source-incidence case, in which the source
field is a spherical wave: $\psi_{s}=e^{ik\rho}/\rho,$ where $\rho
\equiv\left\Vert \vec{r}-\vec{r}_{s}\right\Vert $ is the distance measured
from $\vec{r}$ to the position of the point source $\vec{r}_{s}$. As before,
if $\vec{r}$ happens to lie on the edge of the aperture, then we denote $\rho$
by $\rho_{s},$ and use $\hat{\rho}_{s}$ to denote the unit vector of $\vec
{r}-\vec{r}_{s}.$ As expected, the illuminated region in this case is an
oblique cone with vertex at $\vec{r}_{s},$ as shown in Figure (2).%

{\parbox[b]{3.2413in}{\begin{center}
\includegraphics[
height=1.7348in,
width=3.2413in
]%
{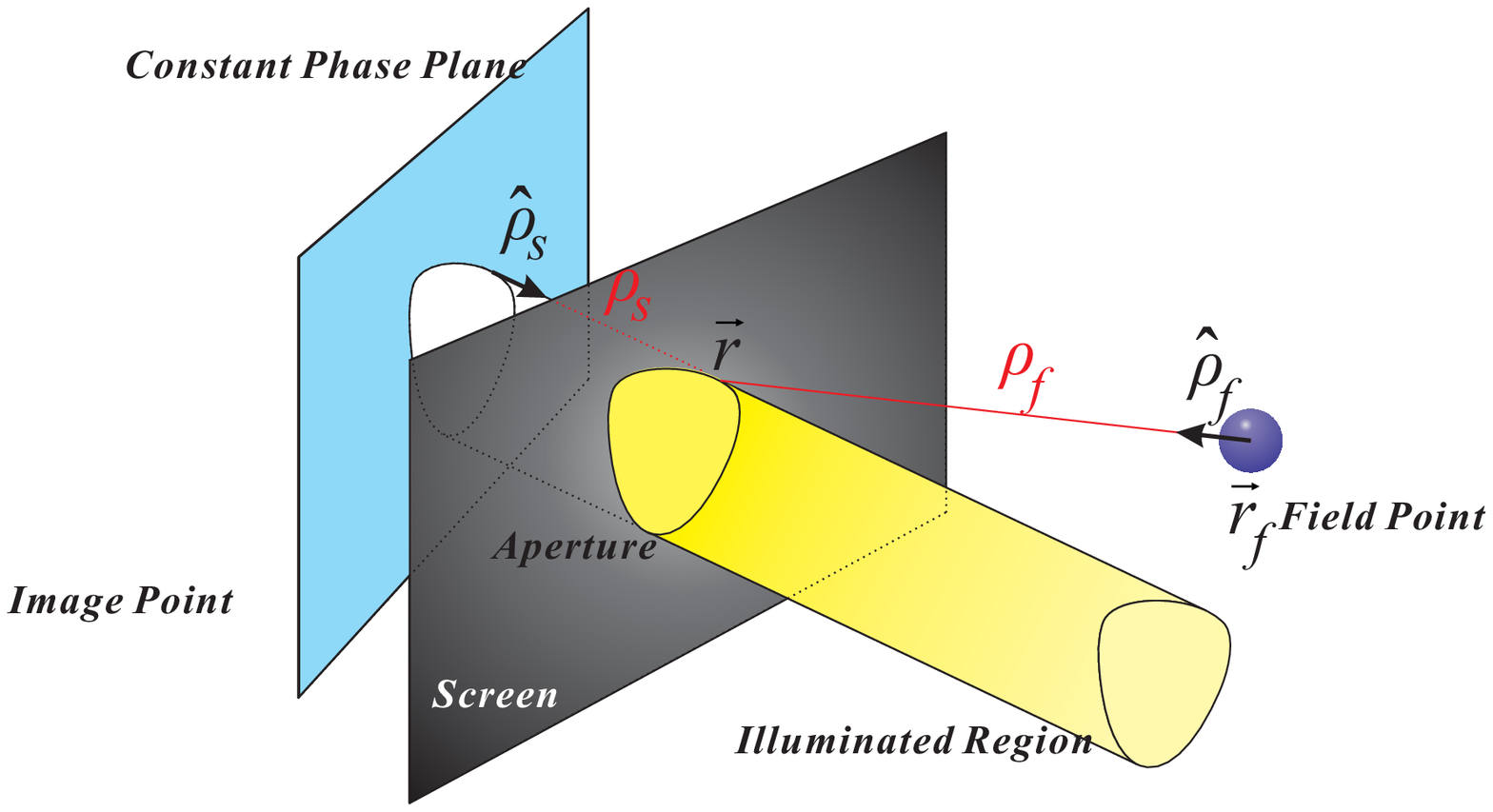}%
\\
Figure (1)
\end{center}}}
{\parbox[b]{3.2603in}{\begin{center}
\includegraphics[
height=1.9865in,
width=3.2603in
]%
{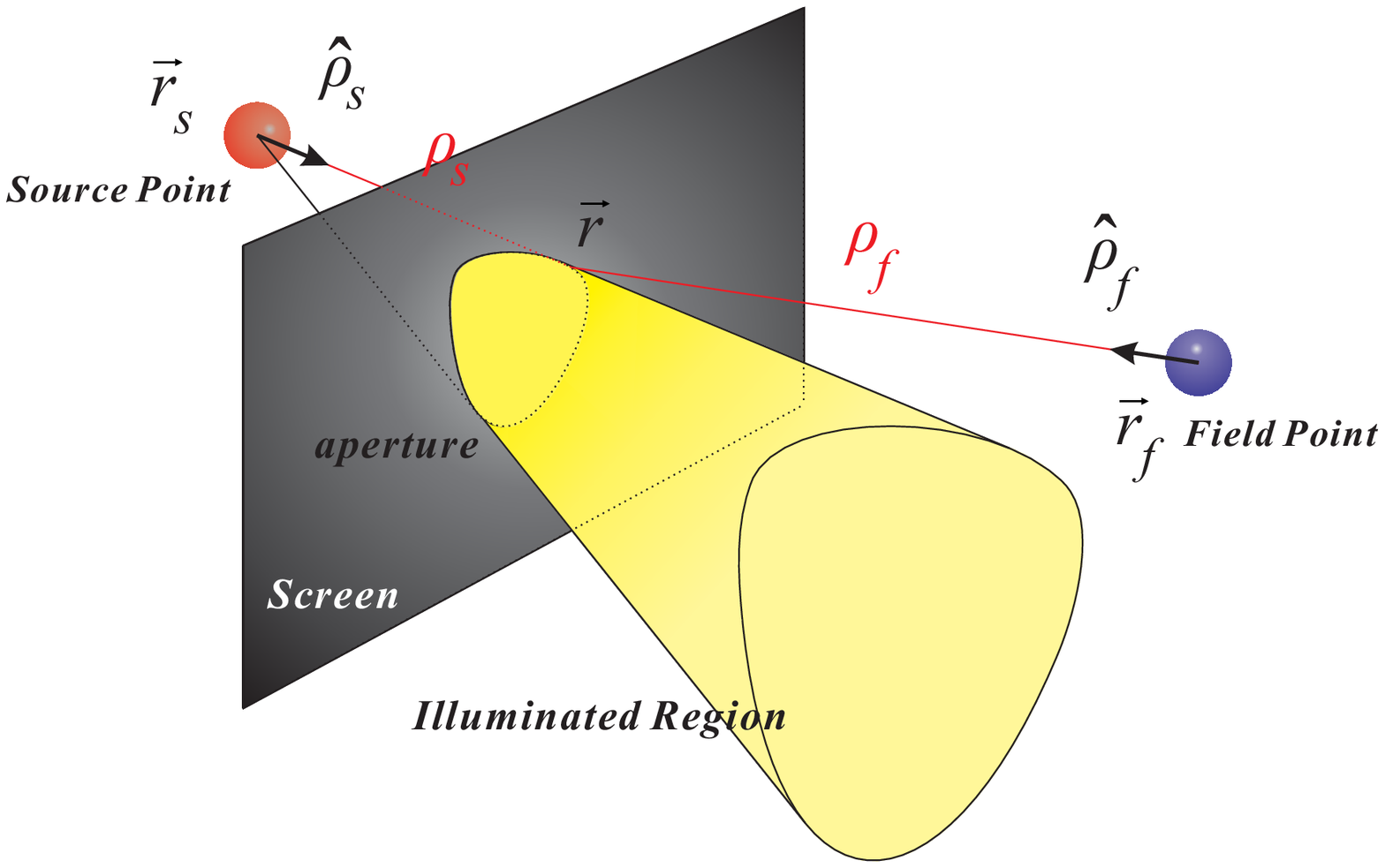}%
\\
Figure (2)
\end{center}}}

\begin{center}

\end{center}

This decomposition realized Young's interpretation for the diffraction
phenomena: Young had once proposed that part of the incident field, which is
called the reflected field, undergone a kind of \textit{reflection} at the
edge of the aperture, and the rest of the incident field, called geometrical
field, just \textit{went through} the aperture without any perturbation, and,
the final diffraction wave was the interference of the two fields
\cite{Sommerfeld Young 311-312}.

But as mentioned earlier, Equation (\ref{psi Kirchhoff}) was derived based on
inconsistent boundary conditions. Now, with the B.C.s proposed by Sommerfeld
(Equation (\ref{BC consistent})), we are able to modify Equation
(\ref{Psi Rubinowicz}) and obtain another similar expression which is not only
mathematically-consistent, but also preserves Young's "field-splitting" interpretation.

Furthermore, Equation (\ref{Psi Rubinowicz}) is not entirely satisfactory in
the sense that the two separated fields are not continuous at the boundary of
the illuminated region. This problem had been discussed by John. S Asvestas,
and he also gave an elegant "solid-angle" representation of $\psi$ which
successfully avoided such discontinuity. However, Asvestas' work was also
based on Kirchhoff's formula, and therefore the mathematical inconsistency
still exists. Besides that, the solid-angle representation derived by Asvestas
does not reduce to electrostatic case in the long wavelength limit
$k\rightarrow0.$ In this paper, we will give a slightly different
decomposition formula with consistent boundary conditions and generalize the
electrostatic result to the diffraction problem.

Another unsatisfactory feature of Rubinowicz' decomposition formula Equation
(\ref{Psi Rubinowicz}) is that the functional form of the line integral is not
what one would expect an ordinary reflection wave should be, due to some
confusing factors in the integrand. And therefore in this paper, we will also
derive a neater representation of the line integral that mimics the behavior
of ordinary reflection in geometrical optics.

In the last section, we'll explain why the boundary conditions (Equation
(\ref{BC consistent})) is more suitable by compare the result of our work with
Sommerfeld's 2-D straight-edge diffraction problem, which is the one of the
few examples where the Helmholtz equation has an exact solution. We'll see
that Kirchhoff's solution has a relative deviation from the exact solution.

\section{Modified Expression for Rubinowicz' Decomposition Formula}

With the boundary conditions Equation (\ref{BC consistent}), we begin from
Sommerfeld's integral formula Equation (\ref{psi Sommerfeld}). Since
$G_{D}\left(  \partial\psi_{s}/\partial n\right)  =0$ on the aperture, we add
it back into the integrand and rearrange a little:%
\begin{align}
\psi\left(  \vec{r}_{f}\right)   &  =-\int_{\text{aperture}}\left(  \psi
_{s}\frac{\partial G_{D}}{\partial n}-G_{D}\frac{\partial\psi_{s}}{\partial
n}\right)  da\nonumber\\
&  =-\int_{\text{aperture}}\left(  \psi_{s}\frac{\partial G_{K}}{\partial
n}-G_{K}\frac{\partial\psi_{s}}{\partial n}\right)  da+\int_{\text{aperture}%
}\left(  \psi_{s}\frac{\partial G_{K}^{\ast}}{\partial n}-G_{K}^{\ast}%
\frac{\partial\psi_{s}}{\partial n}\right)  da, \label{Split GK and GD}%
\end{align}
and in the last line we identify the first integral is nothing but Equation
(\ref{psi Kirchhoff}) and thus equal to Equation (\ref{Psi Rubinowicz}). The
second integral has exactly the same functional form as the first one, except
for the replacement $\vec{r}_{f}\rightarrow\vec{r}_{f}^{\ast}.$ So the final
result is%
\begin{align}
\psi\left(  \vec{r}_{f}\right)   &  =\left\{
\begin{array}
[c]{cl}%
\psi_{s}\left(  \vec{r}_{f}\right)  & ,\vec{r}_{f}\in\text{illuminated
region}\\
0 & ,\text{otherwise}%
\end{array}
\right. \nonumber\\
&  -\frac{1}{4\pi}%
{\displaystyle\oint\nolimits_{\text{edge}}}
\psi_{s}\left(  \vec{r}\right)  \frac{e^{ik\rho_{f}}}{\rho_{f}}\left(
\frac{\hat{\rho}_{s}\times\hat{\rho}_{f}}{1+\hat{\rho}_{s}\cdot\hat{\rho}_{f}%
}-\frac{\hat{\rho}_{s}\times\hat{\rho}_{f}^{\ast}}{1+\hat{\rho}_{s}\cdot
\hat{\rho}_{f}^{\ast}}\right)  \cdot d\vec{l}. \label{Psi Complete}%
\end{align}
(Note that $\vec{r}_{f}^{\ast}$ is on the opposite side of the screen, so it
always lies outside of the illuminated region, and thus there is no
corresponding geometrical field.) Here $\hat{\rho}_{f}^{\ast}$ is the unit
vector of $\vec{r}-\vec{r}_{f}^{\ast}.$ The image term $G_{K}^{\ast}$
contributes another line integral to the final expression, which now seems
more ugly. Actually, Rubinowicz had also derived this result in his paper in
1917. However, for some reason he seemed to abandon this result and used
Equation (\ref{Psi Rubinowicz}) in his successive papers. In the following
sections, we'll show that, with some deformation, Equation (\ref{Psi Complete}%
) (or equivalently, Equation (\ref{psi Sommerfeld})) can take another form
which has some merits mentioned in the introduction.

\section{"Reflective" Representation}

\subsection{Motivations from Geometrical Optics}

In ordinary geometrical optics, the reflection phenomena can be comprehended
this way: given a source distribution, one draw the "image source" behind the
"mirror," as shown in Figure (3), where there is a point real source labeled
by $S$, and the reflected field is equal to the incident field from the image
point $S^{\ast}$.

\bigskip\
{\parbox[b]{3.2811in}{\begin{center}
\includegraphics[
height=2.1594in,
width=3.2811in
]%
{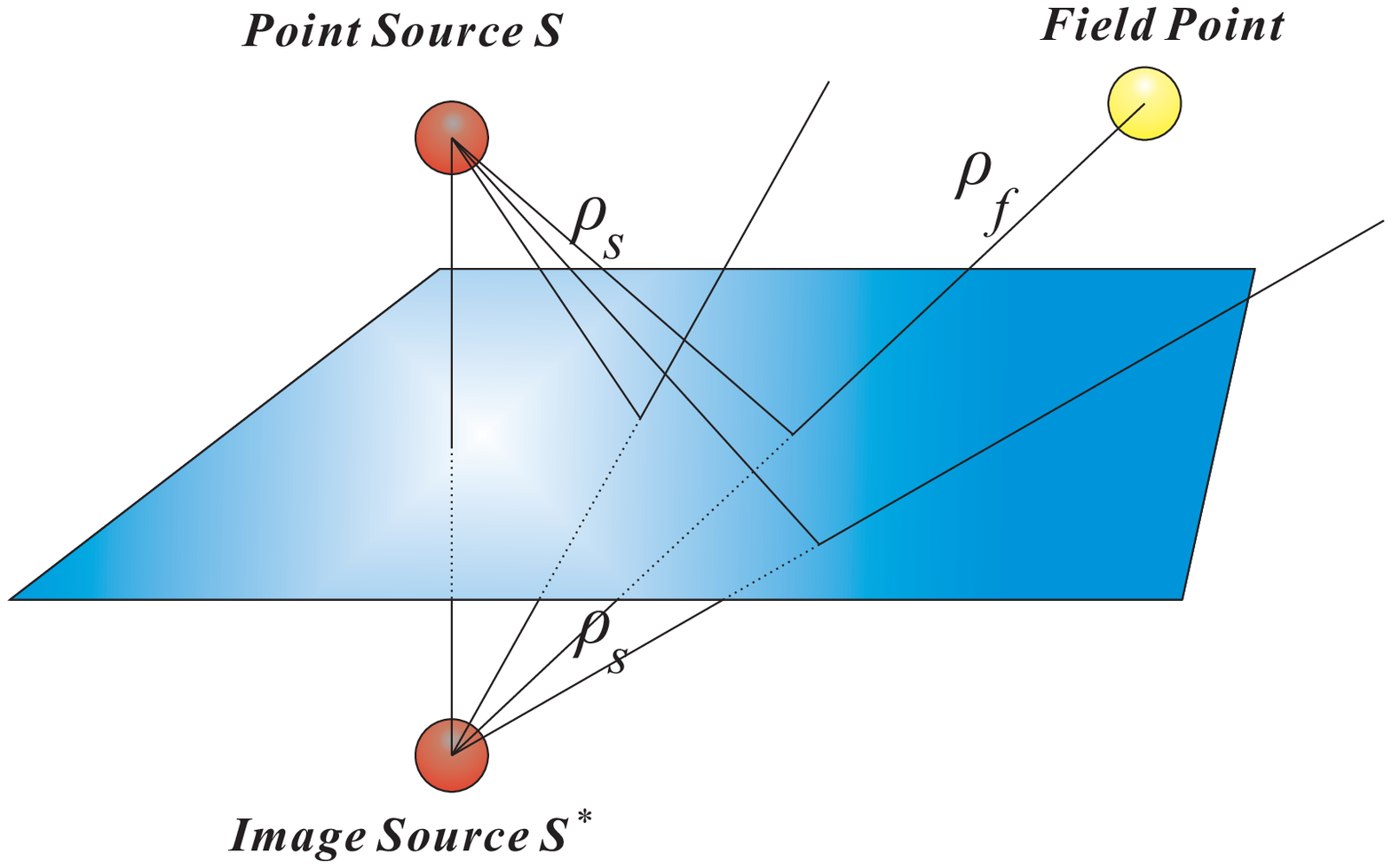}%
\\
Figure (3)
\end{center}}}
{\parbox[b]{2.8323in}{\begin{center}
\includegraphics[
height=2.6809in,
width=2.8323in
]%
{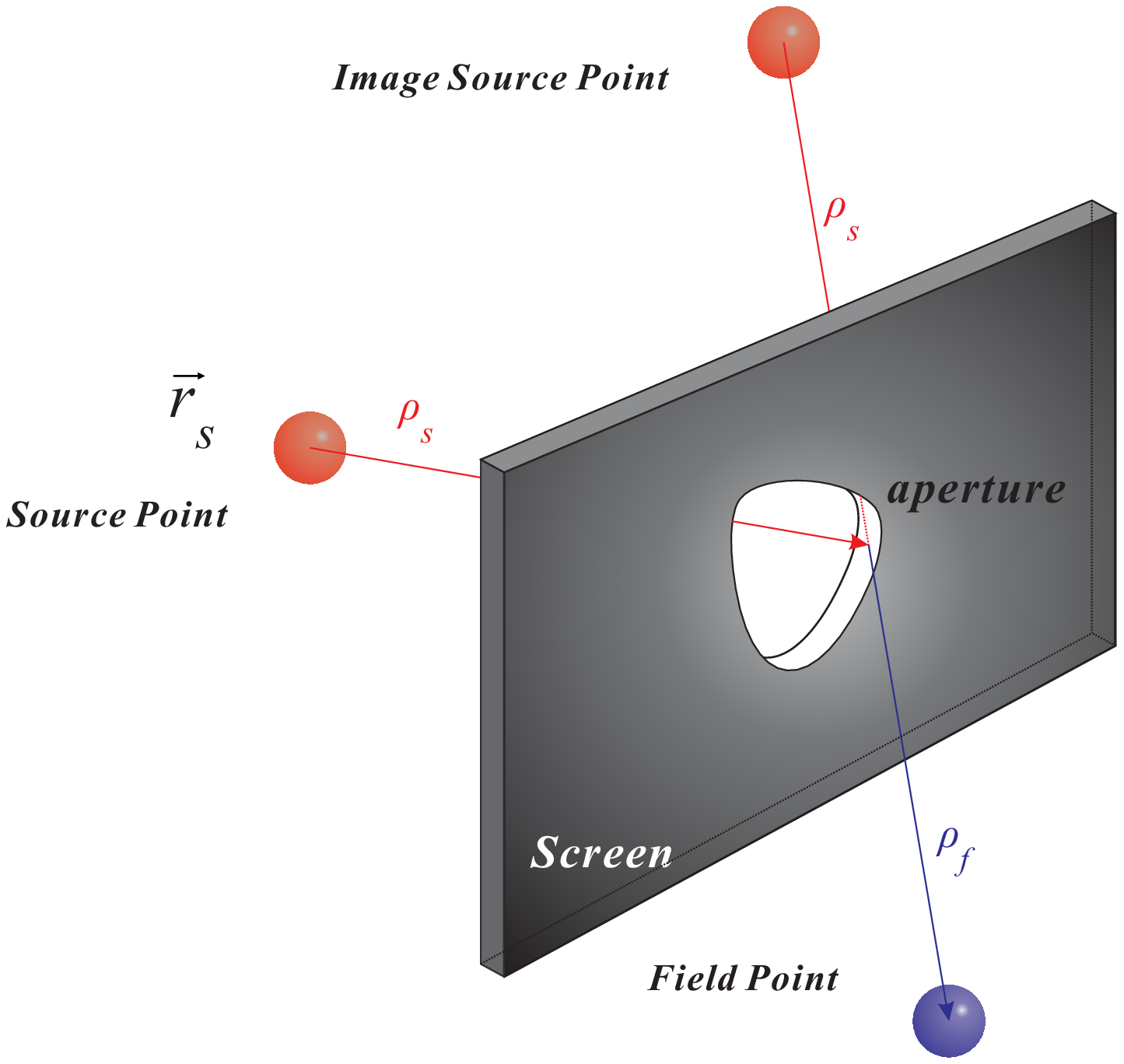}%
\\
Figure (4)
\end{center}}}

\begin{center}

\end{center}

So, if what Young really meant (in the early days when he saw the diffraction
phenomena) by "reflection at the edge" was the reflection in geometrical
optics, then we \textit{expect} the line integral should take the form%
\begin{align}
\psi_{\text{reflection}}  &  \sim%
{\displaystyle\oint_{\text{edge}}}
e^{ik\left(  \rho_{s}+\rho_{f}\right)  }\text{ for plane wave,}%
\label{Expected Reflective Plane}\\
\psi_{\text{reflection}}  &  \sim%
{\displaystyle\oint_{\text{edge}}}
\dfrac{e^{ik\left(  \rho_{s}+\rho_{f}\right)  }}{\rho_{s}+\rho_{f}}\text{ for
point source.} \label{Expected Reflective Point}%
\end{align}

Namely, as shown in Figure (4), we imagine that the screen has a finite
thickness, and as the incident field $\psi_{s}$ comes in, it is reflected by
the "cut" around the aperture, and thus when the reflected field reaches the
field point $\vec{r}_{f},$ the incident field $\psi_{s}$ has propagated for a
total optical length $\rho_{s}+\rho_{f},$ and therefore $\psi
_{\text{reflection}}$ should take the form as Equation
(\ref{Expected Reflective Plane}) or (\ref{Expected Reflective Point}).

These expectations actually can be accomplished by some deformations of
Equation (\ref{Psi Complete}), but let's do it another way: to derive the
$\psi_{\text{reflection}}$ from the beginning Equation (\ref{psi Sommerfeld}),
and this will make derivation more neater.

\subsection{Reflection at the Boundary}

Simplify Equation (\ref{psi Sommerfeld}) a little, and we get%
\begin{align}
\psi\left(  \vec{r}_{f}\right)   &  =-\int_{\text{aperture}}\psi_{s}%
\frac{\partial G_{D}}{\partial n}da=-2\int_{\text{aperture}}\psi_{s}%
\frac{\partial G_{K}}{\partial n}da\nonumber\\
&  =\frac{1}{2\pi}\int_{\text{aperture}}\psi_{s}\frac{\partial}{\partial
r}\left(  \frac{e^{ikr}}{r}\right)  \frac{\partial r}{\partial z}da
\label{Psi before reflective-rep.}%
\end{align}
where $r\equiv\left\Vert \vec{r}-\vec{r}_{f}\right\Vert ,$ and let $\hat
{e}_{z}$ be the inward unit normal to the screen ($\hat{e}_{z}=-\hat{n}$). To
simplify the result, we consider the following two cases:

\textbf{Case 1.} \textit{Plane Wave Diffraction}%

\begin{center}
\includegraphics[
height=2.6005in,
width=3.7023in
]%
{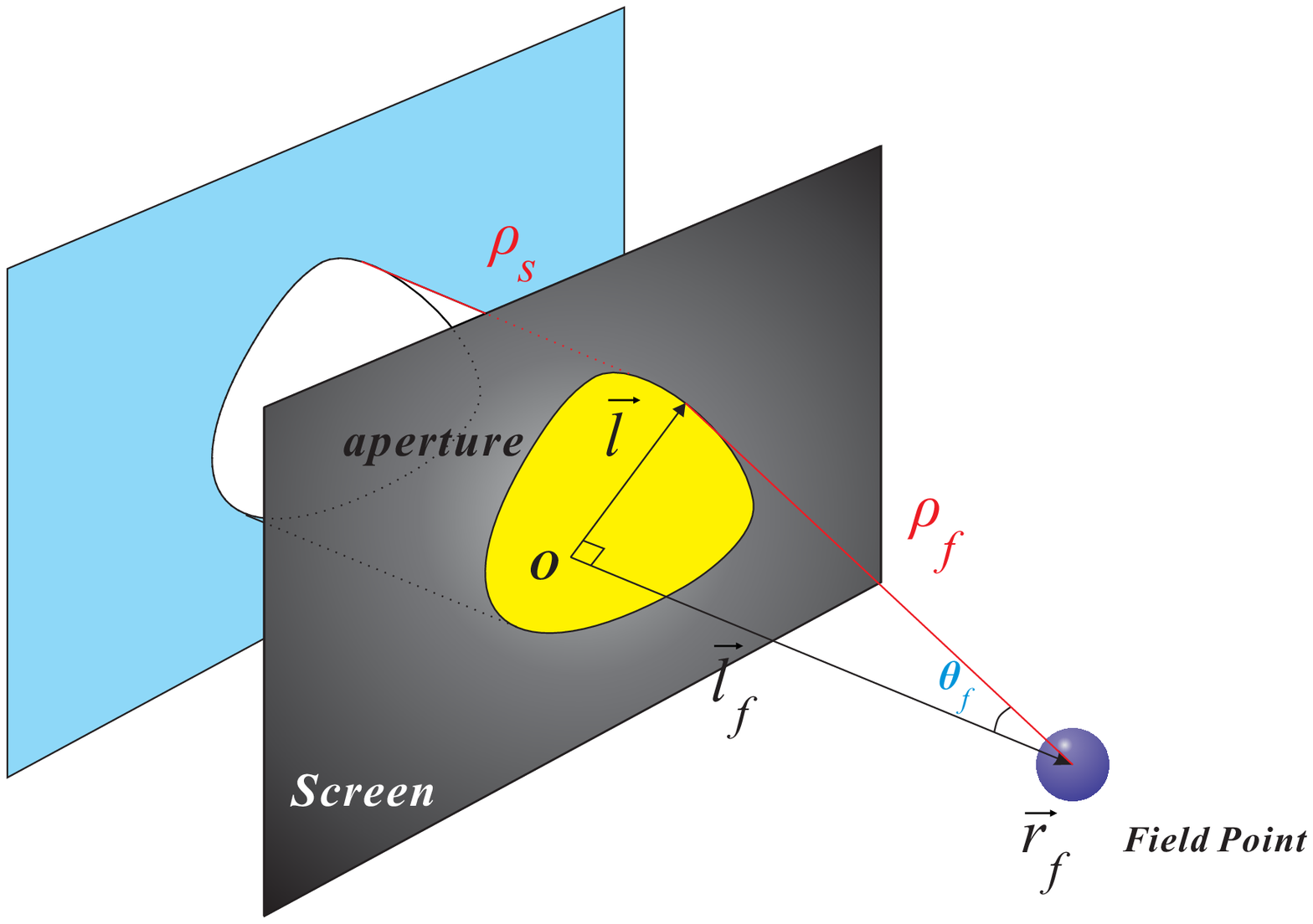}%
\\
Figure (5)
\end{center}

We assume the wave propagates in the direction perpendicular to the
screen--and it is reasonable to make this assumption since experimentally it
is the most common configuration. Under this postulation, $\psi_{s}%
=e^{ik\rho_{s}}$ where $\rho_{s}$ is now a constant quantity representing the
distance from a constant phase plane to the screen. As shown in Figure (5), we
make the projection of the field point $\vec{r}_{f}$ on the screen, and denote
it by $O$. Notice that $O$ does not necessarily lie inside the aperture. Next,
define two vectors $\vec{l}$ and $\vec{l}_{f}$ as shown in Figure (5). Then
every point along $\vec{l}$ can be described by $s\vec{l}$, where $0\leq
s\leq1.$ Therefore%
\[
r\equiv\left\Vert \vec{r}-\vec{r}_{f}\right\Vert =\left\Vert s\vec{l}-\vec
{l}_{f}\right\Vert =\sqrt{s^{2}l^{2}+l_{f}^{2}}%
\]
where $l$ and $l_{f}$ represent the magnitude of $\vec{l}$ and $\vec{l}_{f},$
respectively. The area element on the aperture is%
\[
d\vec{a}=\vec{l}ds\times sd\vec{l},
\]
and Equation (\ref{Psi before reflective-rep.}) can be evaluated as%
\begin{align*}
\psi\left(  \vec{r}_{f}\right)   &  =-\frac{1}{2\pi}\int_{\text{aperture}%
}e^{ik\rho_{s}}\left(  \frac{ik}{r}-\frac{1}{r^{2}}\right)  e^{ikr}\frac
{l_{f}}{r}da\\
&  =-\frac{l_{f}e^{ik\rho_{s}}}{2\pi}%
{\displaystyle\oint\nolimits_{\text{edge}}}
\int_{s=0}^{s=1}\left(  \frac{ik}{s^{2}l^{2}+l_{f}^{2}}-\frac{1}{\left(
s^{2}l^{2}+l_{f}^{2}\right)  ^{3/2}}\right)  e^{ik\sqrt{s^{2}l^{2}+l_{f}^{2}}%
}\left(  \vec{l}ds\times sd\vec{l}\right)  \cdot\hat{e}_{z}.\\
&  =-\frac{l_{f}e^{ik\rho_{s}}}{2\pi}%
{\displaystyle\oint\nolimits_{\text{edge}}}
\frac{\left(  \vec{l}\times d\vec{l}\right)  \cdot\hat{e}_{z}}{l^{2}}\left(
\frac{e^{ik\sqrt{s^{2}l^{2}+l_{f}^{2}}}}{\sqrt{s^{2}l^{2}+l_{f}^{2}}}\right)
\bigg|_{s=0}^{s=1}%
\end{align*}
But%
\[
\left(  \vec{l}\times d\vec{l}\right)  \cdot\hat{e}_{z}=l^{2}d\phi,
\]
where $\phi$ is the angle subtended by the arc of the boundary of the aperture
as measured from $O.$ So%
\[
\psi\left(  \vec{r}_{f}\right)  =\frac{l_{f}e^{ik\rho_{s}}}{2\pi}%
{\displaystyle\oint\nolimits_{\text{edge}}}
d\phi\left(  \frac{e^{ikl_{f}}}{l_{f}}-\frac{e^{ik\sqrt{l^{2}+l_{f}^{2}}}%
}{\sqrt{l^{2}+l_{f}^{2}}}\right)  =\frac{1}{2\pi}%
{\displaystyle\oint\nolimits_{\text{edge}}}
d\phi\left(  \psi_{s}\left(  \vec{r}_{f}\right)  -e^{ik\left(  \rho_{s}%
+\rho_{f}\right)  }\cos\theta_{f}\right)
\]
where $\theta_{f}$ is the angle indicated in Figure (5).

To separate the geometrical and reflected fields apart, we perform the line
integral to the first term of the integrand:
\[
\frac{1}{2\pi}%
{\displaystyle\oint\nolimits_{\text{edge}}}
\psi_{s}\left(  \vec{r}_{f}\right)  d\phi=\left\{
\begin{array}
[c]{cl}%
\psi_{s}\left(  \vec{r}_{f}\right)  & ,\text{if }O\text{ lies inside the
aperture}\\
0 & ,\text{otherwise}%
\end{array}
\right.  .
\]
Since $O$ lies inside the aperture if and only if $\vec{r}_{f}\in$illuminated
region, so%
\begin{equation}
\psi\left(  \vec{r}_{f}\right)  =\left\{
\begin{array}
[c]{cl}%
\psi_{s}\left(  \vec{r}_{f}\right)  & ,\vec{r}_{f}\in\text{illuminated
region}\\
0 & ,\text{otherwise}%
\end{array}
\right.  -\dfrac{1}{2\pi}%
{\displaystyle\oint\nolimits_{\text{edge}}}
e^{ik\left(  \rho_{s}+\rho_{f}\right)  }\cos\theta_{f}d\phi
\label{Psi in reflective form (Plane)}%
\end{equation}

\newpage

\textbf{Case 2.} \textit{Point Source Diffraction}%

\begin{center}
\includegraphics[
height=2.3462in,
width=4.2609in
]%
{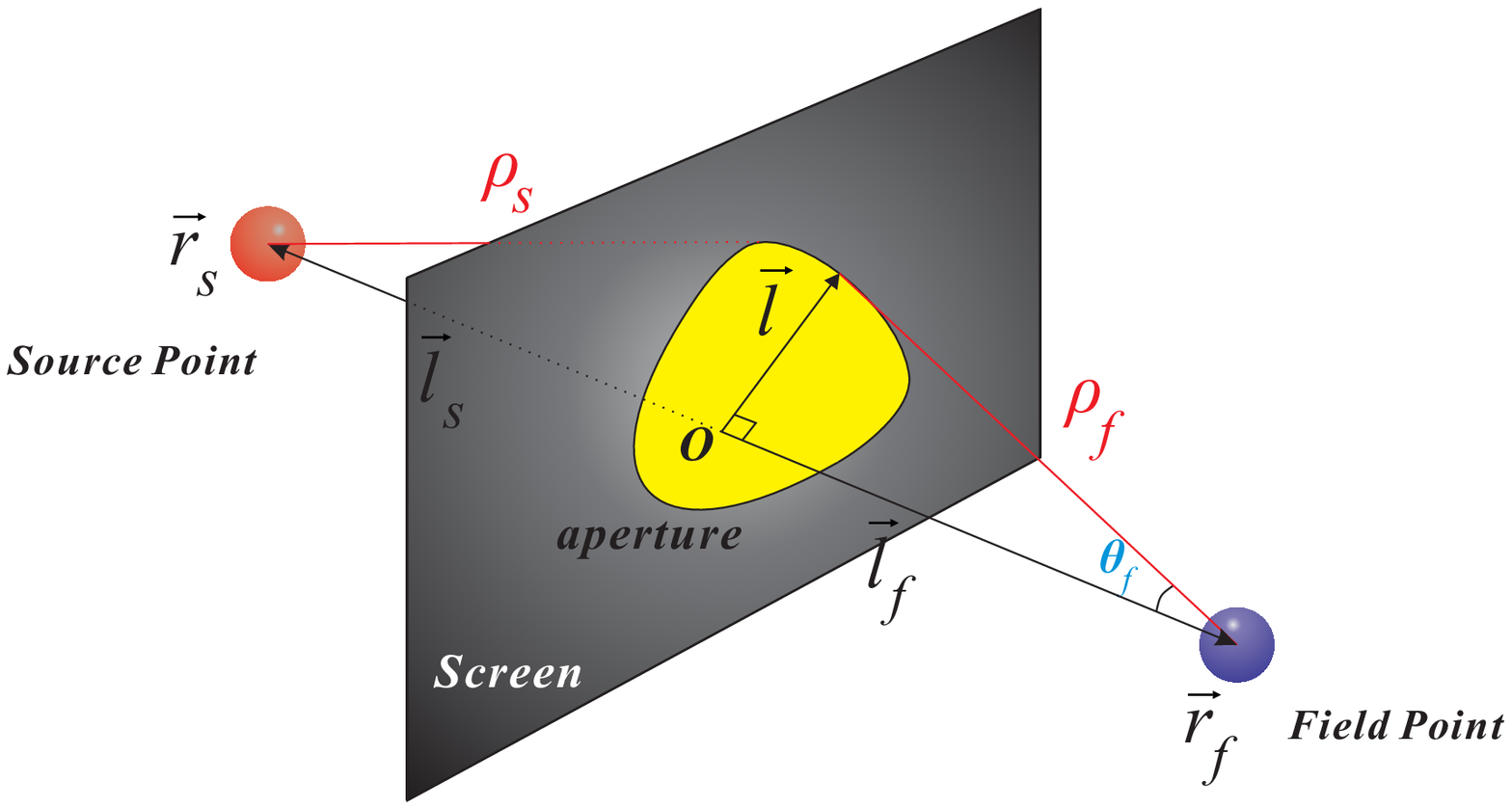}%
\\
Figure (6)
\end{center}

Similar to the previous case, we attempt to assume that $\vec{r}_{f}-\vec
{r}_{s}$ is perpendicular to the screen, that is, we confine $\vec{r}_{f}$ to
lie on the central axis as shown in Figure (6). But this time the assumption
is totally unreasonable--after all we cannot restrict the position of $\vec
{r}_{f}$. We believe that this formulation can be generalized without this
assumption, but for the present we'll consider this special case, and try to
elucidate the idea of "reflection at the boundary."

Identify $\psi_{s}=e^{ik\rho}/\rho,$ where $\rho\equiv\left\Vert \vec{r}%
-\vec{r}_{s}\right\Vert ,$ and define $O$, $\vec{l}_{f},$ $\vec{l}$ as before.
Here, we define a new vector $\vec{l}_{s},$ to be the vector from $O$ to
$\vec{r}_{s}.$ Therefore%
\[
\rho=\left\Vert s\vec{l}-\vec{l}_{s}\right\Vert =\sqrt{s^{2}l^{2}+l_{s}^{2}},
\]
and Equation.(\ref{Psi before reflective-rep.}) can be evaluated as%
\begin{align*}
\psi\left(  \vec{r}_{f}\right)   &  =-\frac{1}{2\pi}\int_{\text{aperture}%
}\frac{e^{ik\rho}}{\rho}\left(  \frac{ik}{r}-\frac{1}{r^{2}}\right)
e^{ikr}\frac{l_{f}}{r}da\\
&  =-\frac{l_{f}}{2\pi}%
{\displaystyle\oint\nolimits_{\text{edge}}}
\int_{s=0}^{s=1}\frac{e^{ik\left(  \sqrt{s^{2}l^{2}+l_{f}^{2}}+\sqrt
{s^{2}l^{2}+l_{s}^{2}}\right)  }}{\sqrt{s^{2}l^{2}+l_{s}^{2}}}\left(
\frac{ik}{s^{2}l^{2}+l_{f}^{2}}-\frac{1}{\left(  s^{2}l^{2}+l_{f}^{2}\right)
^{3/2}}\right)  \left(  \vec{l}ds\times sd\vec{l}\right)  \cdot\hat{e}_{z}\\
&  =-\frac{l_{f}}{2\pi}%
{\displaystyle\oint\nolimits_{\text{edge}}}
\frac{\left(  \vec{l}\times d\vec{l}\right)  \cdot\hat{e}_{z}}{l^{2}}\left(
\frac{e^{ik\sqrt{s^{2}l^{2}+l_{f}^{2}}}}{\sqrt{s^{2}l^{2}+l_{f}^{2}}}%
\frac{e^{ik\sqrt{s^{2}l^{2}+l_{s}^{2}}}}{\sqrt{s^{2}l^{2}+l_{f}^{2}}%
+\sqrt{s^{2}l^{2}+l_{s}^{2}}}\right)  \bigg|_{s=0}^{s=1}%
\end{align*}
So%
\[
\psi\left(  \vec{r}_{f}\right)  =\frac{1}{2\pi}%
{\displaystyle\oint\nolimits_{\text{edge}}}
\left(  \frac{e^{ik\left(  l_{f}+l_{s}\right)  }}{l_{f}+l_{s}}-\frac
{e^{ik\left(  \rho_{s}+\rho_{f}\right)  }}{\rho_{s}+\rho_{f}}\frac{l_{f}}%
{\rho_{f}}\right)  d\phi=\frac{1}{2\pi}%
{\displaystyle\oint\nolimits_{\text{edge}}}
\left(  \psi_{s}\left(  \vec{r}_{f}\right)  -\frac{e^{ik\left(  \rho_{s}%
+\rho_{f}\right)  }}{\rho_{s}+\rho_{f}}\cos\theta_{f}\right)  d\phi
\]
where $\theta_{f}$ has the same definition as before.

Again the geometrical field can be separated out by the same method, and we
get the final result:
\begin{equation}
\psi\left(  \vec{r}_{f}\right)  =\left\{
\begin{array}
[c]{cl}%
\psi_{s}\left(  \vec{r}_{f}\right)  & ,\vec{r}_{f}\in\text{illuminated
region}\\
0 & ,\text{otherwise}%
\end{array}
\right.  -\dfrac{1}{2\pi}%
{\displaystyle\oint\nolimits_{\text{edge}}}
\dfrac{e^{ik\left(  \rho_{s}+\rho_{f}\right)  }}{\rho_{s}+\rho_{f}}\cos
\theta_{f}d\phi\label{Psi in reflective form (Point)}%
\end{equation}

\section{Solid-Angle Representation}

\subsection{Motivation from Electrostatics}

Although Equation (\ref{Psi Complete}) is a mathematically-consistent
solution, it still exhibits the same problem as what Rubinowicz encountered in
his solution: the geometrical and reflected fields are discontinuous at the
boundary of the illuminated region. To overcome this problem, we seek for the
analogy in electrostatics: Consider a grounded infinite conducting plane with
a finite insulating region $\sigma$ at which the potential is held at a
constant value $V_{0}:$%
\[
V=\left\{
\begin{array}
[c]{cl}%
V_{0} & ,\text{on the insulating region }\sigma\\
0 & ,\text{on the conducting plane}%
\end{array}
\right.  .
\]
Assume there is no other source charge in the half space $z>0,$ and therefore
$\nabla^{2}V=0$ there. This boundary value problem has the solution%
\begin{equation}
V\left(  \vec{r}_{f}\right)  =\frac{\Omega_{f}}{2\pi}V_{0}
\label{Electrostatic result}%
\end{equation}
where $\Omega_{f}$ is the solid angle subtended by the region $\sigma$ as
observed at the field point $\vec{r}_{f}.$

Inspired by the electrostatic result, we attempt a solution for diffraction
problem of the form%
\[
\psi\left(  \vec{r}_{f}\right)  =\frac{\Omega_{f}}{2\pi}\psi_{s}\left(
\vec{r}_{f}\right)  +\text{(a line-integral),}%
\]
namely, we expect the geometrical field to take the similar form of Equation
(\ref{Electrostatic result}), while the reflected field remains a
line-integral around the boundary of the aperture. The advantage of this
formulation is that both the geometrical and reflected fields now vary
continuously, without any jump discontinuity across the boundary of the
illuminated region. Different from Asvestas' work, in this formula, we see
that as the field point $\vec{r}_{f}$ approaches to the aperture, then
$\Omega_{f}\rightarrow2\pi,$ and the geometrical field $\rightarrow\psi_{s}$
while we expect the reflected field to vanish totally. That is, if we reside
on the aperture, we are exposing ourself to the source $\psi_{s}$ without the
influence of the edge.

\subsection{Inverse Cone}

We begin from Equation (\ref{Split GK and GD}), and define%
\begin{align*}
J  &  \equiv-%
{\displaystyle\oint_{\text{aperture}}}
\left(  \psi_{s}\frac{\partial G_{K}}{\partial n}-G_{K}\frac{\partial\psi_{s}%
}{\partial n}\right)  da,\\
J^{\ast}  &  \equiv%
{\displaystyle\oint_{\text{aperture}}}
\left(  \psi_{s}\frac{\partial G_{K}^{\ast}}{\partial n}-G_{K}^{\ast}%
\frac{\partial\psi_{s}}{\partial n}\right)  da.
\end{align*}

\bigskip%
\begin{center}
\includegraphics[
height=3.0217in,
width=3.0493in
]%
{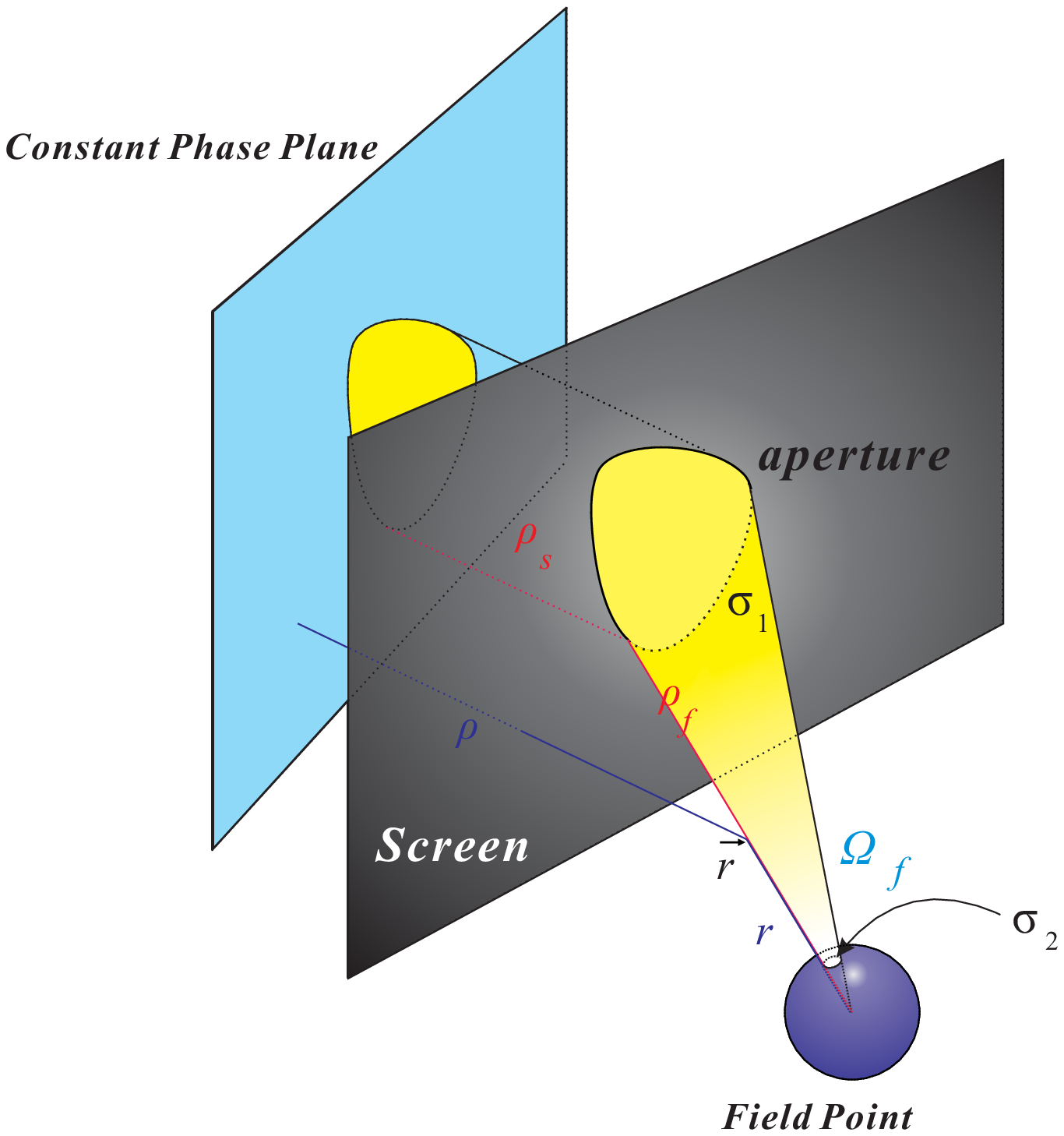}%
\\
Figure (7)
\end{center}

To evaluate $J,$ we do a trick slightly different from what Rubinowicz did. As
shown in Figure (7), we make an auxiliary surface with the vertex at the
\textit{field point}, and make a small ball centered at the field point.
Define $\sigma_{1}$ to be the surface of the cone outside the small ball while
$\sigma_{2}$ to be the surface of the small ball inside the cone. Apply the
divergence theorem to the region enclosed by $\sigma_{1}$, $\sigma_{2}$ and
the aperture:%
\[
\int_{\text{aperture}}\vec{F}\cdot d\vec{a}+\int_{\sigma_{1}}\vec{F}\cdot
d\vec{a}+\int_{\sigma_{2}}\vec{F}\cdot d\vec{a}=0
\]
where $\vec{F}\equiv\psi_{s}\vec{\nabla}G_{K}-G_{K}\vec{\nabla}\psi_{s},$ and
$d\vec{a}$ is the vectorial area element pointing outwardly from the volume
enclosed. If we let the radius of the small ball approach to zero, then%
\[
\int_{\sigma_{2}}\vec{F}\cdot d\vec{a}\rightarrow-\frac{\Omega_{f}}{4\pi
}\left(  \vec{\nabla}\cdot\vec{F}\right)  =-\frac{\Omega_{f}}{4\pi}\psi
_{s}\left(  \vec{r}_{f}\right)  ,
\]
where, as desired, $\Omega_{f}$ is the solid angle subtended by the aperture
as observed at the field point $\vec{r}_{F}.$ So%
\[
J=\frac{\Omega_{f}}{4\pi}\psi_{s}\left(  \vec{r}_{f}\right)  -\int_{\sigma
_{1}}\vec{F}\cdot d\vec{a}.
\]
The surface integral can be evaluated by the same trick presented by
Rubinowicz, as discuss in the following two cases:

\textbf{Case 1.} \textit{Plane Wave Diffraction}

On the auxiliary surface, $\vec{\nabla}G_{K}=0,$ and therefore%
\begin{align*}
\int_{\sigma_{1}}\vec{F}\cdot d\vec{a}  &  =\int_{\sigma_{1}}\frac{e^{ikr}}%
{r}ike^{ik\rho}\hat{\rho}\cdot\left(  \frac{\hat{\rho}_{f}\times d\vec{l}%
}{\rho_{f}}rdr\right)  =\int_{\sigma_{1}}ike^{ik\left(  r+\rho\right)  }%
\hat{\rho}_{s}\cdot\left(  \frac{\hat{\rho}_{f}\times d\vec{l}}{\rho_{f}%
}dr\right) \\
&  =%
{\displaystyle\oint\nolimits_{\text{edge}}}
ik\hat{\rho}_{s}\cdot\left(  \frac{\hat{\rho}_{f}\times d\vec{l}}{\rho_{f}%
}\right)  \int_{r=0}^{r=\rho_{f}}e^{ik\left(  r+\rho\right)  }dr.
\end{align*}

From Figure (7), we have the relation%
\[
\rho=-\left(  \rho_{f}-r\right)  \hat{\rho}_{s}\cdot\hat{\rho}_{f}+\rho_{s}%
\]
and thus%
\[
\int_{r=0}^{r=\rho_{f}}e^{ik\left(  r+\rho\right)  }dr=e^{ik\left(  -\rho
_{f}\hat{\rho}_{s}\cdot\hat{\rho}_{f}+\rho_{s}\right)  }\int_{r=0}^{r=\rho
_{f}}e^{ikr\left(  1+\hat{\rho}_{s}\cdot\hat{\rho}_{f}\right)  }dr=\frac
{1}{ik}\frac{e^{ik\left(  \rho_{s}+\rho_{f}\right)  }-e^{ik\left(  -\rho
_{f}\hat{\rho}_{s}\cdot\hat{\rho}_{f}+\rho_{s}\right)  }}{1+\hat{\rho}%
_{s}\cdot\hat{\rho}_{f}}%
\]
and%
\[
\int_{\sigma_{1}}\vec{F}\cdot d\vec{a}=%
{\displaystyle\oint\nolimits_{\text{edge}}}
e^{ik\rho_{s}}\frac{e^{ik\rho_{f}}}{\rho_{f}}\left(  1-e^{-ik\rho_{f}\left(
1+\hat{\rho}_{s}\cdot\hat{\rho}_{f}\right)  }\right)  \left(  \frac{\hat{\rho
}_{s}\times\hat{\rho}_{f}}{1+\hat{\rho}_{s}\cdot\hat{\rho}_{f}}\right)  \cdot
d\vec{l}.
\]
Therefore%
\[
J=\frac{\Omega_{f}}{4\pi}\psi_{s}\left(  \vec{r}_{f}\right)  -\frac{1}{4\pi}%
{\displaystyle\oint\nolimits_{\text{edge}}}
e^{ik\rho_{s}}\frac{e^{ik\rho_{f}}}{\rho_{f}}\left(  1-e^{-ik\rho_{f}\left(
1+\hat{\rho}_{s}\cdot\hat{\rho}_{f}\right)  }\right)  \left(  \frac{\hat{\rho
}_{s}\times\hat{\rho}_{f}}{1+\hat{\rho}_{s}\cdot\hat{\rho}_{f}}\right)  \cdot
d\vec{l}.
\]

To evaluate $J^{\ast},$ we use the result from Section 2:%
\[
J^{\ast}=\frac{1}{4\pi}%
{\displaystyle\oint\nolimits_{\text{edge}}}
e^{ik\rho_{s}}\frac{e^{ik\rho_{f}}}{\rho_{f}}\left(  \frac{\hat{\rho}%
_{s}\times\hat{\rho}_{f}^{\ast}}{1+\hat{\rho}_{s}\cdot\hat{\rho}_{f}^{\ast}%
}\right)  \cdot d\vec{l}.
\]

Finally, we combine $J$ and $J^{\ast}$:%
\begin{align}
\psi\left(  \vec{r}_{f}\right)   &  =\frac{\Omega_{f}}{4\pi}\psi_{s}\left(
\vec{r}_{f}\right) \nonumber\\
&  -\frac{1}{4\pi}%
{\displaystyle\oint\nolimits_{\text{edge}}}
e^{ik\rho_{s}}\frac{e^{ik\rho_{f}}}{\rho_{f}}\left[  \left(  1-e^{-ik\rho
_{f}\left(  1+\hat{\rho}_{s}\cdot\hat{\rho}_{f}\right)  }\right)  \left(
\frac{\hat{\rho}_{s}\times\hat{\rho}_{f}}{1+\hat{\rho}_{s}\cdot\hat{\rho}_{f}%
}\right)  -\left(  \frac{\hat{\rho}_{s}\times\hat{\rho}_{f}^{\ast}}%
{1+\hat{\rho}_{s}\cdot\hat{\rho}_{f}^{\ast}}\right)  \right]  \cdot d\vec
{l}.\nonumber\\
&  \label{Solid form incomplete (Plane)}%
\end{align}

Although this result fits our demand--$\psi\left(  \vec{r}_{f}\right)  $ is
now expressed in terms of the solid angle $\Omega_{f}$--it is still
unsatisfactory since the denominator of the geometrical part is $4\pi$ instead
of $2\pi.$ Accordingly, if $\vec{r}_{f}$ approaches to the aperture, the
geometrical part only gives us one half of the total source wave $\psi_{s}$,
and thus the reflected part must contribute the rest half part. To fix the
problem, we take the long wavelength limit $k\rightarrow0,$ and thus
\[
\psi_{s}\rightarrow e^{i0\rho}=1,
\]
and Equation (\ref{Solid form incomplete (Plane)}) must be identical to
Equation (\ref{Electrostatic result}):%
\[
\psi\left(  \vec{r}_{f}\right)  \rightarrow\frac{\Omega_{f}}{4\pi}+\frac
{1}{4\pi}%
{\displaystyle\oint\nolimits_{\text{edge}}}
\frac{1}{\rho_{f}}\left(  \frac{\hat{\rho}_{s}\times\hat{\rho}_{f}^{\ast}%
}{1+\hat{\rho}_{s}\cdot\hat{\rho}_{f}^{\ast}}\right)  \cdot d\vec{l}%
\equiv\frac{\Omega_{f}}{2\pi}.
\]
So we have a line-integral representation of solid angle:%
\begin{equation}
\frac{\Omega_{f}}{4\pi}=\frac{1}{4\pi}%
{\displaystyle\oint\nolimits_{\text{edge}}}
\frac{1}{\rho_{f}}\left(  \frac{\hat{\rho}_{s}\times\hat{\rho}_{f}^{\ast}%
}{1+\hat{\rho}_{s}\cdot\hat{\rho}_{f}^{\ast}}\right)  \cdot d\vec{l}.
\label{Solid Angle Formula}%
\end{equation}
This equation has also been derived by Asvestas \cite{Asvestas}, and by
Yih-Yuh Chen \cite{YYC} from a more elegant perspective. \bigskip Note that
since the wave number $k$ vanishes, the vector $\hat{\rho}_{s}$ now can point
in an arbitrary direction, so the representation above is \textit{not unique}.

Finally, we construct the desired $\psi\left(  \vec{r}_{f}\right)  $ by adding
Equation (\ref{Solid Angle Formula}) into Equation
(\ref{Solid form incomplete (Plane)})
\[
\psi\left(  \vec{r}_{f}\right)  =\dfrac{\Omega_{f}}{2\pi}\psi_{s}\left(
\vec{r}_{f}\right)  -\dfrac{1}{4\pi}%
{\displaystyle\oint\nolimits_{\text{edge}}}
\psi_{s}\dfrac{e^{ik\rho_{f}}}{\rho_{f}}\left[
\begin{array}
[c]{c}%
\left(  1-e^{-ik\rho_{f}\left(  1+\hat{\rho}_{s}\cdot\hat{\rho}_{f}\right)
}\right)  \left(  \dfrac{\hat{\rho}_{s}\times\hat{\rho}_{f}}{1+\hat{\rho}%
_{s}\cdot\hat{\rho}_{f}}\right) \\
-\left(  1-e^{-ik\left(  \rho_{s}+\rho_{f}\right)  }\right)  \left(
\dfrac{\hat{\rho}_{s}\times\hat{\rho}_{f}^{\ast}}{1+\hat{\rho}_{s}\cdot
\hat{\rho}_{f}^{\ast}}\right)
\end{array}
\right]  \cdot d\vec{l}.
\]

\textbf{Case 2.} \textit{Point Source Diffraction}

Again, on the auxiliary surface, $\vec{\nabla}G_{K}=0,$ and%

\begin{align*}
\int_{\sigma_{1}}\vec{F}\cdot d\vec{a}  &  =\int_{\sigma_{1}}\frac
{e^{ik\left(  r+\rho\right)  }}{r}\left(  \frac{ik}{\rho}-\frac{1}{\rho^{2}%
}\right)  \hat{\rho}\cdot\left(  \frac{\hat{\rho}_{f}\times d\vec{l}}{\rho
_{f}}rdr\right)  =\int_{\sigma_{1}}e^{ik\left(  r+\rho\right)  }\left(
\frac{ik}{\rho}-\frac{1}{\rho^{2}}\right)  \frac{\vec{\rho}_{s}}{\rho}%
\cdot\left(  \frac{\hat{\rho}_{f}\times d\vec{l}}{\rho_{f}}dr\right) \\
&  =%
{\displaystyle\oint\nolimits_{\text{edge}}}
\vec{\rho}_{s}\cdot\left(  \frac{\hat{\rho}_{f}\times d\vec{l}}{\rho_{f}%
}\right)  \int_{r=0}^{r=\rho_{f}}e^{ik\left(  r+\rho\right)  }\left(
\frac{ik}{\rho^{2}}-\frac{1}{\rho^{3}}\right)  dr
\end{align*}

\bigskip%
\begin{center}
\includegraphics[
height=2.6143in,
width=3.5224in
]%
{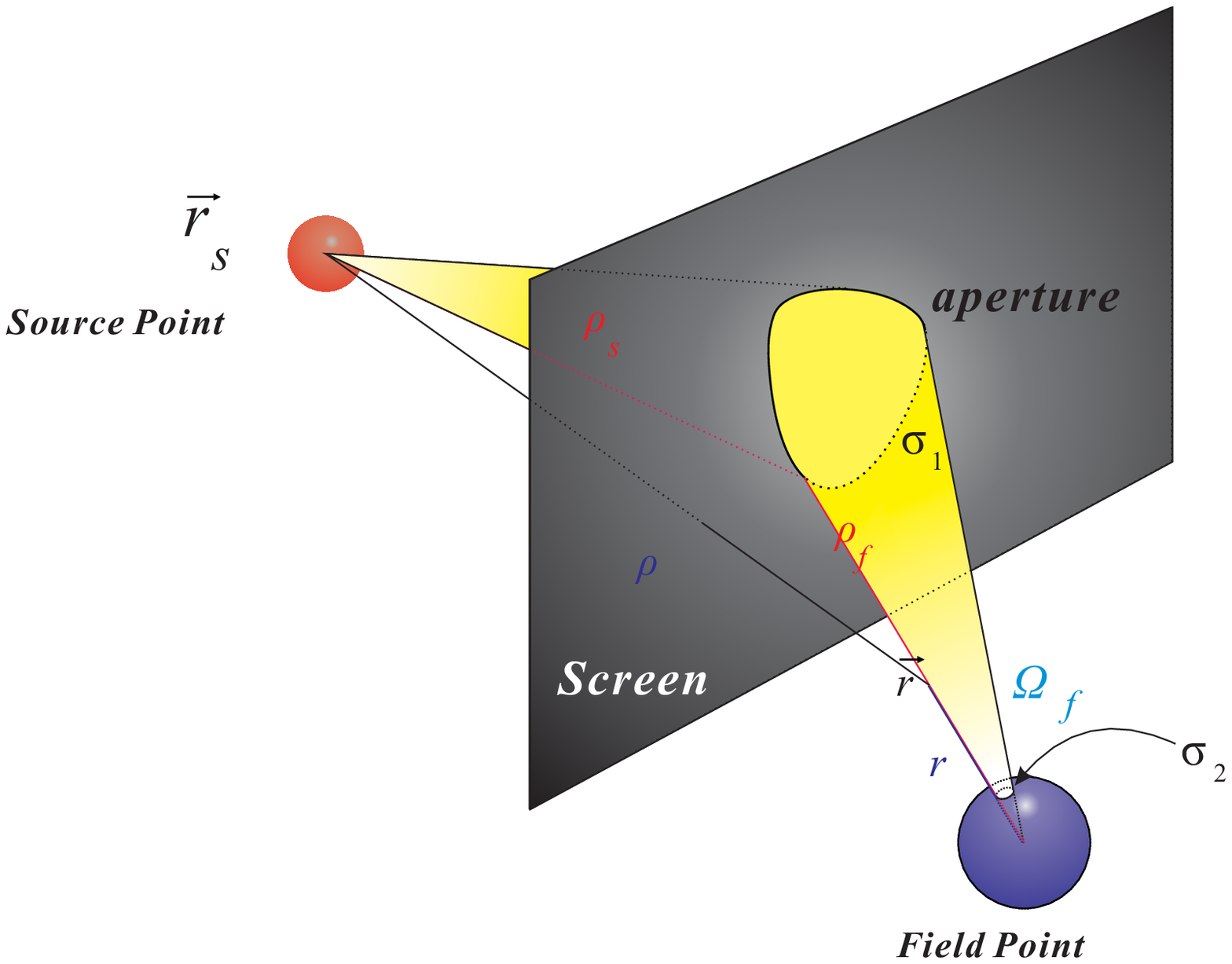}%
\\
Figure (8)
\end{center}

From Figure (8), we have the relation%
\[
\rho^{2}=\rho_{s}^{2}+\left(  \rho_{f}-r\right)  ^{2}-2\rho_{s}\left(
\rho_{f}-r\right)  \hat{\rho}_{s}\cdot\hat{\rho}_{f}.
\]
Differentiate it with respect to $r:$%
\[
\rho\left(  1+\frac{d\rho}{dr}\right)  =r+\rho-\rho_{f}+\rho_{s}\hat{\rho}%
_{s}\cdot\hat{\rho}_{f},
\]
and (follow Rubinowicz' calculation)%
\[
\int_{r=0}^{r=\rho_{f}}e^{ik\left(  r+\rho\right)  }\left(  \frac{ik}{\rho
^{2}}-\frac{1}{\rho^{3}}\right)  dr=\frac{e^{ik\left(  r+\rho\right)  }}%
{\rho\left(  r+\rho-\rho_{f}+\rho_{s}\hat{\rho}_{s}\cdot\hat{\rho}_{f}\right)
}\bigg|_{r=0}^{r=\rho_{f}}=\frac{e^{ik\left(  \rho_{f}+\rho_{s}\right)  }%
}{\rho_{s}^{2}\left(  1+\hat{\rho}_{s}\cdot\hat{\rho}_{f}\right)  }%
-\frac{e^{ik\rho_{0}}}{\rho_{0}^{2}\left(  1+\hat{\rho}_{0}\cdot\hat{\rho}%
_{f}\right)  }%
\]
where $\vec{\rho}_{0}\equiv\vec{\rho}_{s}-\vec{\rho}_{f}$ is the vector from
$\vec{r}_{s}$ to $\vec{r}_{f}.$ So%
\[
\int_{\sigma_{1}}\vec{F}\cdot d\vec{a}=%
{\displaystyle\oint\nolimits_{\text{edge}}}
\frac{e^{ik\rho_{s}}}{\rho_{s}}\frac{e^{ik\rho_{f}}}{\rho_{f}}\left(  \frac
{1}{1+\hat{\rho}_{s}\cdot\hat{\rho}_{f}}-\frac{\rho_{s}^{2}}{\rho_{0}^{2}%
}\frac{e^{-ik\left(  \rho_{s}+\rho_{f}-\rho_{0}\right)  }}{1+\hat{\rho}%
_{0}\cdot\hat{\rho}_{f}}\right)  \left(  \hat{\rho}_{s}\times\hat{\rho}%
_{f}\right)  \cdot d\vec{l}.
\]
Therefore%
\begin{align*}
J  &  =\frac{\Omega_{f}}{4\pi}\psi_{s}\left(  \vec{r}_{f}\right) \\
&  -\frac{1}{4\pi}%
{\displaystyle\oint\nolimits_{\text{edge}}}
\frac{e^{ik\rho_{s}}}{\rho_{s}}\frac{e^{ik\rho_{f}}}{\rho_{f}}\left(  \frac
{1}{1+\hat{\rho}_{s}\cdot\hat{\rho}_{f}}-\frac{\rho_{s}^{2}}{\rho_{0}^{2}%
}\frac{e^{-ik\left(  \rho_{s}+\rho_{f}-\rho_{0}\right)  }}{1+\hat{\rho}%
_{0}\cdot\hat{\rho}_{f}}\right)  \left(  \hat{\rho}_{s}\times\hat{\rho}%
_{f}\right)  \cdot d\vec{l}.
\end{align*}

Again, we use the result from Section 2:%
\[
J^{\ast}=\frac{1}{4\pi}%
{\displaystyle\oint\nolimits_{\text{edge}}}
\frac{e^{ik\rho_{s}}}{\rho_{s}}\frac{e^{ik\rho_{f}}}{\rho_{f}}\left(
\frac{\hat{\rho}_{s}\times\hat{\rho}_{f}^{\ast}}{1+\hat{\rho}_{s}\cdot
\hat{\rho}_{f}^{\ast}}\right)  \cdot d\vec{l}.
\]
Combine $J$ and $J^{\ast},$ we have%
\begin{align}
\psi\left(  \vec{r}_{f}\right)   &  =\frac{\Omega_{f}}{4\pi}\psi_{s}\left(
\vec{r}_{f}\right) \nonumber\\
&  -\frac{1}{4\pi}%
{\displaystyle\oint\nolimits_{\text{edge}}}
\frac{e^{ik\rho_{s}}}{\rho_{s}}\frac{e^{ik\rho_{f}}}{\rho_{f}}\left[
\begin{array}
[c]{c}%
\left(  \dfrac{1}{1+\hat{\rho}_{s}\cdot\hat{\rho}_{f}}-\dfrac{\rho_{s}^{2}%
}{\rho_{0}^{2}}\dfrac{e^{-ik\left(  \rho_{s}+\rho_{f}-\rho_{0}\right)  }%
}{1+\hat{\rho}_{0}\cdot\hat{\rho}_{f}}\right)  \left(  \hat{\rho}_{s}%
\times\hat{\rho}_{f}\right) \\
-\dfrac{\hat{\rho}_{s}\times\hat{\rho}_{f}^{\ast}}{1+\hat{\rho}_{s}\cdot
\hat{\rho}_{f}^{\ast}}%
\end{array}
\right]  \cdot d\vec{l}.\nonumber\\
&  \label{Solid form incomplete}%
\end{align}

To construct the correct factor $1/2\pi,$ we use Equation
(\ref{Solid Angle Formula}) to add another $\left(  \Omega_{f}/4\pi\right)
\psi_{s}$ to the geometric wave. But note that $\hat{\rho}_{s}$ in Equation
(\ref{Solid Angle Formula}) is an arbitrary \textit{constant} vector, and in
Equation (\ref{Solid form incomplete}) $\hat{\rho}_{s}$ represents a varying
vector that changes its direction as we integrate along the edge. Thus we must
specify one direction for $\hat{\rho}_{s}$ in Equation
(\ref{Solid Angle Formula}) so that we can insert it into Equation
(\ref{Solid form incomplete}). The result is most symmetric if we adopt
$\hat{\rho}_{s}\equiv\hat{\rho}_{0}$ in Equation (\ref{Solid Angle Formula})
\[
\psi\left(  \vec{r}_{f}\right)  =\dfrac{\Omega_{f}}{2\pi}\psi_{s}\left(
\vec{r}_{f}\right)  -\dfrac{1}{4\pi}%
{\displaystyle\oint\nolimits_{\text{edge}}}
\psi_{s}\dfrac{e^{ik\rho_{f}}}{\rho_{f}}\left[
\begin{array}
[c]{c}%
\left(  \dfrac{1}{1+\hat{\rho}_{s}\cdot\hat{\rho}_{f}}-\dfrac{\rho_{s}^{2}%
}{\rho_{0}^{2}}\dfrac{e^{-ik\left(  \rho_{s}+\rho_{f}-\rho_{0}\right)  }%
}{1+\hat{\rho}_{0}\cdot\hat{\rho}_{f}}\right)  \left(  \hat{\rho}_{s}%
\times\hat{\rho}_{f}\right) \\
-\dfrac{\hat{\rho}_{s}\times\hat{\rho}_{f}^{\ast}}{1+\hat{\rho}_{s}\cdot
\hat{\rho}_{f}^{\ast}}+e^{-ik\rho_{f}}\left(  \dfrac{\hat{\rho}_{0}\times
\hat{\rho}_{f}^{\ast}}{1+\hat{\rho}_{0}\cdot\hat{\rho}_{f}^{\ast}}\right)
\end{array}
\right]  \cdot d\vec{l}.
\]

We have seen that in both cases the field $\psi\left(  \vec{r}_{f}\right)  $
has the form%
\[
\psi\left(  \vec{r}_{f}\right)  =\dfrac{\Omega_{f}}{2\pi}\psi_{s}\left(
\vec{r}_{f}\right)  -\dfrac{1}{4\pi}%
{\displaystyle\oint\nolimits_{\text{edge}}}
\psi_{s}\dfrac{e^{ik\rho_{f}}}{\rho_{f}}\left[  \cdots\right]  \cdot d\vec
{l},
\]
and it is obvious that both the geometrical and the line integral parts of
$\psi$ are now continuous across the surface of illuminated region. However,
the integrand in $\left[  \cdots\right]  $ now depends on the type of the
source. There is another point to be mentioned: since $\Omega_{f}\propto
1/\rho_{f}{}^{2},$ in the far zone the geometrical field is overwhelmed by the
reflected field, which is proportional to $1/\rho_{f}$ of the source wave. The
situation is reversed in the near zone, of course.

\section{Comparison of Boundary Conditions}

As discussed in previous sections, the boundary conditions based on
Kirchhoff's theory is mathematically inconsistent, and by using the proper
Green's function, the diffraction theory can be transformed into a boundary
value problem of Dirichlet type which is mathematically admissible. However,
this is not the whole story. As the source wave $\psi_{s}$ propagates toward
the aperture, the wave must be modified by the presence of the opaque screen,
and thus $\psi$ is not exactly equal to $\psi_{s}$, the unperturbed source, on
the aperture. So the boundary values Equation (\ref{BC consistent}) imposed
earlier is still, unsatisfactory in the physical sense.

However, Sommerfeld has solved a $2-D$ straight edge diffraction problem
rigorously without using the unperturbed source wave as boundary values
\cite{Sommerfeld pp249}, and we'll see in a moment that, by some deformation
of Equation (\ref{Psi Complete}), the functional form of our solution is very
closed to that of Sommerfeld's, and, therefore, we may regard Equation
(\ref{BC consistent}) as an acceptable approximation to the real, rigorous solution.

\subsection{Approximated Solution for a Point Source}

Consider an infinite half plane lying on $z=0$ and $x>0,$ with a point source
lying in the region $z<0$ as before. The solution $\psi$ in the space $z>0$
can be solved by Equation (\ref{Psi Complete})%
\begin{align*}
\psi\left(  \vec{r}_{f}\right)   &  =\left\{
\begin{array}
[c]{cl}%
\psi_{s}\left(  \vec{r}_{f}\right)  & ,\vec{r}_{f}\text{ }\in\text{illuminated
region}\\
0 & ,\text{otherwise}%
\end{array}
\right. \\
&  -\frac{1}{4\pi}\int_{\text{edge}}A\frac{e^{ik\rho_{s}}}{\rho_{s}}%
\frac{e^{ik\rho_{f}}}{\rho_{f}}\left(  \frac{\hat{\rho}_{s}\times\hat{\rho
}_{f}}{1+\hat{\rho}_{s}\cdot\hat{\rho}_{f}}-\frac{\hat{\rho}_{s}\times
\hat{\rho}_{f}^{\ast}}{1+\hat{\rho}_{s}\cdot\hat{\rho}_{f}^{\ast}}\right)
\cdot d\vec{l}.
\end{align*}
where the line integral is performed along the infinite straight
edge\footnote{And thus the integral sign $\int$ is used instead of $%
{\textstyle\oint}
.$}. The amplitude $A$ of the source field is now expressed explicitly for
later convenience.

In the far field region $k\rho_{f}\gg1,$ we apply stationary-phase
approximation to evaluate the reflected field:%
\begin{equation}
I=\frac{1}{4\pi}\int_{\text{edge}}A\frac{e^{ik\left(  \rho_{s}+\rho
_{f}\right)  }}{\rho_{s}\rho_{f}}\left(  \frac{\hat{\rho}_{s}\times\hat{\rho
}_{f}}{1+\hat{\rho}_{s}\cdot\hat{\rho}_{f}}-\frac{\hat{\rho}_{s}\times
\hat{\rho}_{f}^{\ast}}{1+\hat{\rho}_{s}\cdot\hat{\rho}_{f}^{\ast}}\right)
\cdot d\vec{l}. \label{stationary app before}%
\end{equation}
The the stationary point occurs when $\nabla\left(  \rho_{s}+\rho_{f}\right)
\cdot d\vec{l}=0,$ and we expand the phase at the stationary point:%
\begin{align}
\rho_{s}+\rho_{f}  &  =\left(  \rho_{s}+\rho_{f}\right)  \bigg|_{0}%
+\nabla\left(  \rho_{s}+\rho_{f}\right)  \bigg|_{0}\cdot\delta\vec{l}+\frac
{1}{2}\delta\vec{l}\cdot\left(  \frac{\mathbf{\hat{1}}-\hat{\rho}_{s}\hat
{\rho}_{s}}{\rho_{s}}+\frac{\mathbf{\hat{1}}-\hat{\rho}_{f}\hat{\rho}_{f}%
}{\rho_{f}}\right)  \bigg|_{0}\cdot\delta\vec{l}\nonumber\\
&  =\left(  \rho_{s}+\rho_{f}\right)  \bigg|_{0}+\frac{1}{2}\frac{\rho
_{s}+\rho_{f}}{\rho_{s}\rho_{f}}\delta\vec{l}^{2}\sin^{2}\left(  d\vec{l}%
,\hat{\rho}_{s}\right)  \bigg|_{0}, \label{phase app}%
\end{align}
where the \bigskip subscript $0$ denotes the stationary point, which in this
special case is the point on the edge nearest to $\vec{r}_{f}$; also,
$\mathbf{\hat{1}}$ is the identity operator in three dimensional space, and
$\sin\left(  d\vec{l},\hat{\rho}_{s}\right)  $ is the sine of the angle
between $d\vec{l}$ and $\hat{\rho}_{s}.$ Insert Equation (\ref{phase app})
into Equation (\ref{stationary app before}), and perform the Gaussian
integral, we get
\[
I\simeq\frac{1}{4\pi}\frac{Ae^{ik\left(  \rho_{s}+\rho_{f}\right)  }}{\rho
_{s}\rho_{f}}\left(  \frac{\hat{\rho}_{s}\times\hat{\rho}_{f}}{1+\hat{\rho
}_{s}\cdot\hat{\rho}_{f}}-\frac{\hat{\rho}_{s}\times\hat{\rho}_{f}^{\ast}%
}{1+\hat{\rho}_{s}\cdot\hat{\rho}_{f}^{\ast}}\right)  \cdot\frac{d\vec{l}%
}{\left\Vert d\vec{l}\right\Vert }\bigg|_{0}\sqrt{\frac{2\pi i}{k}\frac
{\rho_{s}\rho_{f}}{\rho_{s}+\rho_{f}}}\frac{1}{\sin\left(  d\vec{l},\hat{\rho
}_{s}\right)  }\bigg|_{0}.
\]
But%
\[
\left(  \frac{\hat{\rho}_{s}\times\hat{\rho}_{f}}{1+\hat{\rho}_{s}\cdot
\hat{\rho}_{f}}-\frac{\hat{\rho}_{s}\times\hat{\rho}_{f}^{\ast}}{1+\hat{\rho
}_{s}\cdot\hat{\rho}_{f}^{\ast}}\right)  \cdot\frac{d\vec{l}}{\left\Vert
d\vec{l}\right\Vert }\bigg|_{0}=\left(  \frac{\hat{\rho}_{f}\cdot\hat{n}%
}{1+\hat{\rho}_{s}\cdot\hat{\rho}_{f}}-\frac{\hat{\rho}_{f}^{\ast}\cdot\hat
{n}}{1+\hat{\rho}_{s}\cdot\hat{\rho}_{f}^{\ast}}\right)  \sin\left(  d\vec
{l},\hat{\rho}_{s}\right)  \bigg|_{0}.
\]
where
\[
\hat{n}=\frac{d\vec{l}\times\hat{\rho}_{s}}{\left\Vert d\vec{l}\times\hat
{\rho}_{s}\right\Vert }%
\]
is the unit outward normal of the geometric light cone. To simplify the factor
in the parenthesis, refer to Figure (9), we have%
\[
\frac{\hat{\rho}_{f}\cdot\hat{n}}{1+\hat{\rho}_{s}\cdot\hat{\rho}_{f}}%
-\frac{\hat{\rho}_{f}^{\ast}\cdot\hat{n}}{1+\hat{\rho}_{s}\cdot\hat{\rho}%
_{f}^{\ast}}=\frac{-\sin\left(  \phi-\alpha\right)  }{1+\cos\left(
\phi-\alpha\right)  }-\frac{\sin\left(  \phi+\alpha\right)  }{1+\cos\left(
\phi+\alpha\right)  }=-\frac{2\sin\phi}{\cos\alpha+\cos\phi}%
\]

\bigskip%
\begin{center}
\includegraphics[
height=2.4146in,
width=2.8824in
]%
{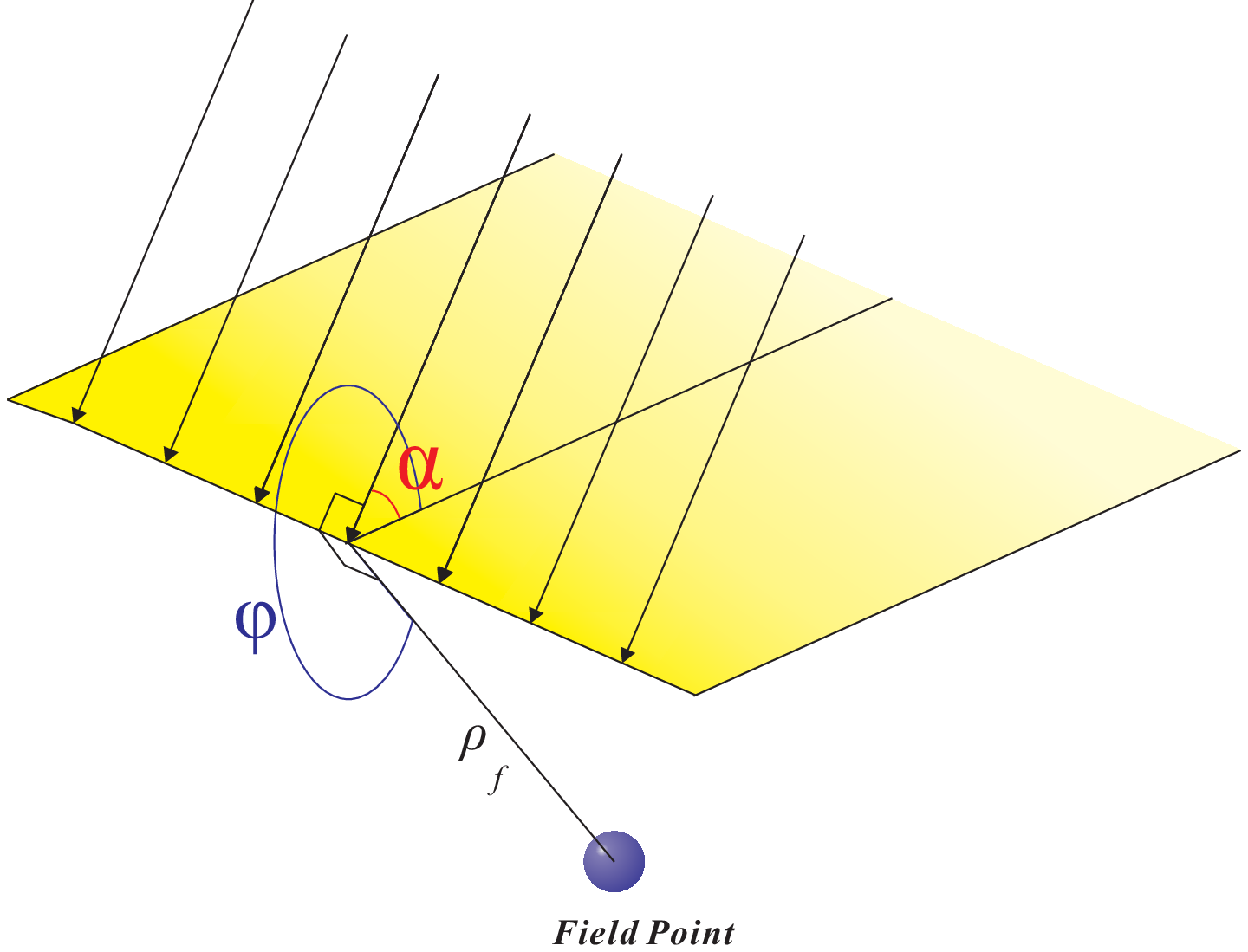}%
\\
Figure (9)
\end{center}

So finally,%
\begin{equation}
I\simeq-A\frac{e^{ik\left(  \rho_{s}+\rho_{f}\right)  }}{\sqrt{2\pi k\rho
_{s}\rho_{f}}}\frac{2\sin\phi}{\cos\alpha+\cos\phi}\frac{e^{i\frac{\pi}{4}}%
}{\sqrt{\rho_{s}+\rho_{f}}} \label{Appx Sol Point}%
\end{equation}
The factor $e^{i\frac{\pi}{4}}$ can explain the reason why the diffraction
pattern in the water has a phase delay compared to the incident wave.

\subsection{Approximated Solution for Plane Waves}

In the case of plane-wave-incidence, we can obtain the solution from Equation
(\ref{Appx Sol Point}) by taking the limit%
\[
\rho_{s}\rightarrow\infty,\text{ }A\rightarrow\infty,\text{ while keeping
}\frac{A}{\rho_{s}}\rightarrow\text{finite number, taken to be }1
\]
The result is%

\[
I\rightarrow-\frac{e^{ik\left(  \rho_{s}+\rho_{f}\right)  }}{\sqrt{2\pi
k\rho_{f}}}\left(  \frac{2\sin\phi}{\cos\alpha+\cos\phi}\right)  e^{i\frac
{\pi}{4}}=-\frac{1+i}{4\sqrt{\pi k\rho_{f}}}e^{ik\rho_{f}}\left(  \dfrac
{1}{\cos\frac{\phi-\alpha}{2}}-\dfrac{1}{\cos\frac{\phi+\alpha}{2}}\right)
\frac{\cos\frac{\phi}{2}}{\sin\frac{\alpha}{2}}.
\]

Here $e^{ik\rho_{s}}$ has been dropped since Sommerfeld assumed that the plane
wave has phase $0$ right at $\rho_{f}=0$.

So the total field is
\[
\psi\left(  \vec{r}_{f}\right)  \simeq\left\{
\begin{array}
[c]{rl}%
\psi_{s}\left(  \vec{r}_{f}\right)  +\dfrac{1+i}{4\sqrt{\pi k\rho_{f}}%
}e^{ik\rho_{f}}\left(  \dfrac{1}{\cos\frac{\phi-\alpha}{2}}-\dfrac{1}%
{\cos\frac{\phi+\alpha}{2}}\right)  \dfrac{\cos\frac{\phi}{2}}{\sin
\frac{\alpha}{2}} & ,\vec{r}_{f}\in\text{illuminated region}\\
\dfrac{1+i}{4\sqrt{\pi k\rho_{f}}}e^{ik\rho_{f}}\left(  \dfrac{1}{\cos
\frac{\phi-\alpha}{2}}-\dfrac{1}{\cos\frac{\phi+\alpha}{2}}\right)
\dfrac{\cos\frac{\phi}{2}}{\sin\frac{\alpha}{2}} & ,\text{otherwise}%
\end{array}
\right.  .
\]
In comparison with Sommerfeld's solution \cite{Sommerfeld pp249}%

\[
\psi\left(  \vec{r}_{f}\right)  \simeq\left\{
\begin{array}
[c]{rl}%
\psi_{s}\left(  \vec{r}_{f}\right)  +\dfrac{1+i}{4\sqrt{\pi k\rho_{f}}%
}e^{ik\rho_{f}}\left(  \dfrac{1}{\cos\frac{\phi-\alpha}{2}}-\dfrac{1}%
{\cos\frac{\phi+\alpha}{2}}\right)  & ,\vec{r}_{f}\in\text{illuminated
region}\\
\dfrac{1+i}{4\sqrt{\pi k\rho_{f}}}e^{ik\rho_{f}}\left(  \dfrac{1}{\cos
\frac{\phi-\alpha}{2}}-\dfrac{1}{\cos\frac{\phi+\alpha}{2}}\right)  &
,\text{otherwise}%
\end{array}
\right.  .
\]
we see that, apart from the factor $\dfrac{\cos\frac{\phi}{2}}{\sin
\frac{\alpha}{2}}$, the two representations are similar. The discrepancy
results from the different boundary conditions as discussed before.

If, however, we use the Kirchhoff's integral formula (with mathematically
inconsistent B.C.s), we would obtain%

\[
\psi\left(  \vec{r}_{f}\right)  \simeq\left\{
\begin{array}
[c]{rl}%
\psi_{s}\left(  \vec{r}_{f}\right)  +\dfrac{1+i}{4\sqrt{\pi k\rho_{f}}%
}e^{ik\rho_{f}}\tan\left(  \dfrac{\phi-\alpha}{2}\right)  & ,\vec{r}_{f}%
\in\text{illuminated region}\\
\dfrac{1+i}{4\sqrt{\pi k\rho_{f}}}e^{ik\rho_{f}}\tan\left(  \dfrac{\phi
-\alpha}{2}\right)  & ,\text{otherwise}%
\end{array}
\right.  ,
\]
which has a different functional form from Sommerfeld's solution.

\section{Conclusion}

By using mathematically consistent boundary conditions, we have seen that
Rubinowicz' decomposition formulation can be more useful: the functional form
of the line integral becomes much neater and admits a simple interpretation of
\textit{reflection at edges}. The formulation also provides us a different
approach that makes the diffraction phenomena similar to the electrostatic
problem by using solid angle representation for the geometrical field.
Finally, the diffracted field predicted by this formulation is much closer to
the physical solution, as discussed in the last section.


\begin{thebibliography}{9}                                                                                                %


\bibitem {Rubinowicz Original}V. A. Rubinowicz, Ann. Phys. 53, 257 (1917).

\bibitem {Sommerfeld Young 311-312}Arnold. Sommerfeld, \textit{Lectures on
Theoretical Physics }\textbf{Vol}\textit{.}(IV)\textit{\ Optics}, p.311-312

\bibitem {Asvestas}J. S. Asvestas, J. Opt. Soc. Am. \textbf{2}, No.6, 896 (1985).

\bibitem {YYC}Yih-Yuh Chen, Chinese Journal of Physcis, \textbf{48}, No.3, 324

\bibitem {Sommerfeld pp249}P. 249 of \cite{Sommerfeld Young 311-312}.

\bibitem {Born Wolf}M. Born \& E. Wolf, \textit{Principles of Optics}, 3rd ed.
(Cambridge University Press, Cambridge, New York, 1999)\bigskip
\end{thebibliography}
\end{document}